\documentstyle[twocolumn,prl,aps,psfig]{revtex}
\tightenlines
\begin{document}
\draft
\title{Spin switching in semiconductor quantum dots through spin-orbit 
coupling}
\author{Manuel Val\'{\i}n-Rodr\'{\i}guez, Antonio Puente,
and Lloren\c{c} Serra}
\address{Departament de F\'{\i}sica, Universitat de les Illes Balears,
E-07071 Palma de Mallorca, Spain}
\author{Enrico Lipparini}
\address{Dipartimento di Fisica, Universit\`a di Trento,
and INFM sezione di Trento, I-38050 Povo, Italy}
\date{March 28, 2002}
\maketitle
\begin{abstract}
The spin-orbit coupling influences the total spin of semiconductor
quantum dots. We analyze the theoretical prediction for the 
combined effects of spin-orbit coupling, weak vertical magnetic fields 
and deformation of the dot. Our results allow the characterization of 
the quantum dots as spin switches, controllable with electric gates.
\end{abstract}


A novel technology based on the use of the electron spin, as opposed 
to the more traditional use of the electron charge, is emerging
under the name of spintronics. Several spin-based electronic 
devices have already proved their importance, even at a commercial 
level, such as the spin-valve read heads. A review of the status of
this incipient field can be found in Ref.\ \cite{Wol01}.

The spin-orbit (SO) coupling is an essential mechanism for most 
spintronic devices, since it links the spin and the charge dynamics, 
opening the possibility of spin control through 
electric fields \cite{Dat90}.
Indeed, recent experimental and theoretical investigations 
have shown that the SO coupling  
affects the charge transport and, more specifically, the  
conductance fluctuations of chaotic quantum dots in a
parallel magnetic field \cite{Fol01,Hal01,Ale01}. It also 
affects the dot far-infrared absorption, introducing peculiar 
correlations between the charge and spin oscillating 
densities \cite{Val02}.
In this work we analyze the combined effects of SO coupling, weak 
vertical magnetic ($B$) fields, and spatial deformation in fixing 
the spin and other ground state properties of model semiconductor
dots. 

It will be shown that a sufficiently strong SO coupling can lead 
to spin inversion, with an alternating $B$-dependence, similar
to the observations from capacitance spectroscopy experiments 
of both vertical \cite{Ash96} and lateral quantum 
dots \cite{Cio00}. 
Since SO coupling is also active at low $B$'s, this
mechanism influences the weak field regime and can provide an
alternative interpretation to the one of
Ciorga et al.\ \cite{Cio00} based on calculations for
high magnetic fields.

We consider the SO coupling terms as reviewed by
Voskoboynikov {\em et al.} \cite{Vos01}. Assuming a 2D system
and the effective Hamiltonian formalism, the relevant Dresselhaus
contribution for the standard (001) plane of GaAs reads
\begin{equation}
\label{eq1}
{\cal H}_D = \frac{\lambda_D}{\hbar} \sum_{i=1}^{N}{ 
\left[\, P_x\sigma_x-P_y\sigma_y\,\right]_i} \; ,
\end{equation}  
where the $\sigma$'s are the Pauli matrices and 
${\bf P}=-i\hbar\nabla+\frac{e}{c}{\bf A}$ represents the canonical 
momentum given in terms of the vector potential ${\bf A}$ 
\cite{gauge}.
The Dresselhaus parameter $\lambda_D$ is determined by the dot vertical 
width $z_0$ as \cite{Vos01} $\lambda_D\approx\gamma(\pi/z_0)^2$,
with a  material-specific constant $\gamma$ that for GaAs is
$\gamma=27.5$~eV{\AA$^3$} \cite{Kna96}. A SO coupling of Rashba 
type \cite{Vos01} was also considered, although its contribution to 
the results for GaAs shown below turned out to be negligible.

Neglecting for the moment the Coulomb interaction the complete 
Hamiltonian reads 
${\cal H}={\cal H}_0+{\cal H}_D+{\cal H}_Z$,
where ${\cal H}_0$ consists of the kinetic and confinement 
energies, i.e.,
\begin{equation}
\label{eq2}
{\cal H}_0 = \sum_{i=1}^N{\left[\frac{{\bf P}^2}{2m}+
\frac{1}{2}m(\omega_x^2 x^2+\omega_y^2 y^2)\right]_i}\; .
\end{equation}
Note that the assumed anisotropic confinement will permit 
the modelling of dots with varying elliptical shapes. 
The Zeeman term ${\cal H}_Z$ depends on the total vertical 
spin $S_z$,
the Bohr magneton $\mu_B$ and the effective gyromagnetic factor 
$g^*$, which for bulk GaAs is $-0.44$. Namely, 
${\cal H}_Z = g^* \mu_B B S_z$.

Assuming ${\cal H}_0 \gg {\cal H}_D \gg {\cal H}_Z $ and 
expanding in powers of $\lambda_D$ an analytic diagonalization 
to $O(\lambda_D^3)$ in spin space is possible with 
a unitary transformation \cite{Ale01} 
$\tilde{\cal H}= U_1^+{\cal H}U_1$, giving the transformed Hamiltonian 
\begin{eqnarray}
\label{eq3}
\tilde{\cal H} &=& \sum_{j=1}^N{ 
\left[\rule{0cm}{0.5cm}\right.
{{\bf P}^2\over 2m} 
+ \frac{1}{2}m(\omega_x^2 x^2+\omega_y^2 y^2)}
+ \lambda_D^2 \frac{m}{\hbar^3}\, L_z \sigma_z \nonumber\\
&+& \frac{1}{2}\, g^* \mu_B B \sigma_z
\left.\rule{0cm}{0.5cm}\right]_j 
-N \lambda_D^2\frac{m}{\hbar^2}
+ O(\lambda_D^3)\; ,
\end{eqnarray}
where we have defined the {\em canonical} angular 
momentum operator $L_z=xP_y-yP_x$. 
Despite the spin diagonalization, the $x$ and $y$ degrees of freedom 
in Eq.\ (\ref{eq3}) are still  coupled
through the vector potential in the kinetic energy and in $L_z$.
With a second transformation each spin component can be recast 
in a separable form using the methods of Meyer, 
Kucar and Cederbaum \cite{Mey88}. Specifically, defining 
$\hat{\cal H}_\eta=U_{2\eta}^+\tilde{\cal H}_\eta U_{2\eta}$, with 
$\eta=\uparrow,\downarrow$, we obtain
\begin{eqnarray}
\label{eqS}
\hat{\cal H}_\eta &=& 
\sum_{j=1}^{N_\eta}{ \left[\rule{0cm}{0.5cm}\right. }
\frac{p_x^2}{2M_{1\eta}}+\frac{M_{1\eta}}{2}\Omega_{1\eta}^2 x^2
+ \frac{p_y^2}{2M_{2\eta}}+\frac{M_{2\eta}}{2}\Omega_{2\eta}^2 y^2\nonumber\\
&+& \frac{1}{2}\, g^*\mu_B B s_\eta
\left.\rule{0cm}{0.5cm}\right]_j 
-N_\eta \lambda_D^2\frac{m}{\hbar^2}
+ O(\lambda_D^3)\; ,
\end{eqnarray}
where $s_\eta=\pm 1$ for $\eta=\uparrow,\downarrow$. Assuming,
without loss of generality, $\omega_x\ge\omega_y$ 
the masses and frequencies of the decoupled oscillators are 
\begin{eqnarray}
\label{eqm}
M_{k\eta} &=&
{2m\sqrt{\left(\omega_x^2+\omega_y^2+\omega_{c\eta}^2\right)^2
-4\omega_x^2\omega_y^2}\over
\omega_x^2-\omega_y^2\pm\omega_{c\eta}^2
+\sqrt{\left(\omega_x^2+\omega_y^2+\omega_{c\eta}^2\right)^2
-4\omega_x^2\omega_y^2}}\nonumber\\
\Omega_{k\eta} &=&
{1\over \sqrt{2}}
\left(\rule{0cm}{0.5cm}\right.
{\omega_x^2+\omega_y^2+\omega_{c\eta}^2}\nonumber\\
& \pm & 
\sqrt{\left(\omega_x^2+\omega_y^2+\omega_{c\eta}^2\right)^2
-4\omega_x^2\omega_y^2}
\left.\rule{0cm}{0.5cm}\right)^{1/2}\; ,
\end{eqnarray}
with the upper (lower) sign in $\pm$ corresponding to $k=1$(2).
We have defined in Eq.\ (\ref{eqm}) a spin dependent 
cyclotron frequency
including the SO correction
\begin{equation}
\omega_{c\uparrow,\downarrow}={eB\over mc}
\pm\, 2\lambda_D^2 \frac{m}{\hbar^3}\; .
\end{equation}
The solution to Eq.\ (\ref{eqS}) is given by products of $x$ and $y$ harmonic
oscillator functions which, when transformed back to the laboratory
frame, yield the desired solutions to the original Hamiltonian 
of Eq.\ (\ref{eq3}). The eigenvalues for each spin can be labelled by the number 
of quanta in the $x$ and $y$ oscillators:
\begin{eqnarray}
\label{eqsp}
\varepsilon_{N_1N_2\eta} &=&
\left( N_1+\frac{1}{2} \right) \hbar\Omega_{1\eta} +
\left( N_2+\frac{1}{2} \right) \hbar\Omega_{2\eta} \nonumber\\
&+& s_\eta \frac{1}{2} g^* \mu_B B 
-\lambda_D^2\frac{m}{\hbar^2}\; .
\end{eqnarray}

Figure 1 displays the magnetic field evolution of the single-particle
energies (\ref{eqsp}) for a circular dot,
except for a constant 
representing the charging energy. These values give the chemical 
potential of the dot for varying electron number, as measured for 
instance in capacitance spectroscopy
experiments \cite{Ash96,Cio00}. 
Actually the parabola coefficient 
has been taken from a fit to the experiments \cite{Ash93}. The spin 
is indicated in Fig.\ 1 with the light and dark gray tones. 
In the absence of SO coupling each line corresponds to a given spin,
except for a very small fluctuation due to the Zeeman energy
at some cusps and valleys. In this case the traces arrange themselves in 
parallel pairs of up and down spin. As shown in the two lower panels
of Fig.\ 1, the SO coupling produces sizeable up and down 
fluctuations of the spin. These spin inversions are due to the 
level rearrangements embodied in Eq.\ \ref{eqsp}.  
For $z_0=100$ \AA, i.e., weak SO coupling, the fluctuations start at low
magnetic fields and they extend up to $B\approx$ 1 T. 
An even stronger SO ($z_0=60$ \AA) produces spin inversions up to the
last level crossing, which marks the filling factor $\nu=2$ 
line \cite{Ash93}. Besides, in the latter case the traces are no longer 
paired but, instead, anticorrelated with a $\pi$ phase shift; specially
in the region just before $\nu=2$. 
The results of Fig.\ 1 can help to interpret the experiments 
of Ref.\ \cite{Ash93} and Ref.\ \cite{Cio00}, which observed
anticorrelated behavior in the traces and spin alternation 
with increasing $B$, respectively.

Having analyzed the non-interacting model we shall next estimate 
Coulomb interaction effects by including the selfconsistent Hartree
potential
\begin{equation}
V_H({\bf r})= \frac{e^2}{\kappa}
\int{d{\bf r}' {\rho({\bf r}')\over |{\bf r}'-{\bf r}|}}\; ,
\end{equation}
where $\kappa$ is the semiconductor dielectric constant ($\kappa=12.4$ for
GaAs). Electronic exchange and correlation energies will be considered 
using 
the local-spin density approximation (LSDA) within a spinorial
formalism, since ${\cal H}_D$ breaks the symmetry of a single spin 
quantization axis. The LSDA relies on Monte Carlo calculations for the 
non polarized and fully polarized electron gases at $B=0$ \cite{Tan89}.
A functional theory including current density dependence, 
better adapted to systems in a magnetic field, 
is known to exist \cite{Vig88}.
Nevertheless, current density corrections are quite small
for moderate magnetic fields and, besides, the numerical
solution of the current-density-functional equations is almost unfeasible
in a symmetry unrestricted case, 
unless strong smoothing approximations are introduced \cite{Man}. 
In the results shown below we consider weak vertical magnetic fields, 
as quantified by the condition $\nu\ge 8$, where the filling
factor $\nu$ is 
obtained from the dot central density $\rho_C$ as 
$\nu=2\pi\ell_B^2\rho_C$ (with $\ell_B=\hbar c/eB$ the magnetic length).

Defining the spin density matrix in terms of the Kohn-Sham
spinors $\{{\varphi}_i({\bf r},\eta),i=1,\dots,N\}$ as
\begin{equation}
\rho_{\eta\eta'}({\bf r}) = \sum_{i=1}^N{
\varphi_i^*({\bf r},\eta)\,
\varphi_i({\bf r},\eta')}\; ,
\end{equation}
the LSDA exchange-correlation functional 
$E_{xc}[\rho_{\eta\eta'}]$
yields the following $2\times2$ potential matrix 
\begin{equation}
\label{Vxc}
V_{{\em xc},\eta\eta'}({\bf r}) =
{\delta{E_{xc}}[\rho_{\eta\eta'}]\over
\delta\rho_{\eta\eta'}({\bf r})}\; .
\end{equation}
Details on the evaluation of the functional derivatives in LSDA 
can be found in Ref.\ \cite{Hei99}. The resulting Kohn-Sham 
equations read
\begin{eqnarray}
\label{KS}
\left[\rule{0cm}{0.5cm}\right.
{{\bf P}^2\over 2m} &+&
\frac{1}{2}m\left( \omega_x^2 x^2 +\omega_y^2 y^2\right)
+ s_\eta \frac{1}{2}g^*\mu_B B \nonumber\\
&&
\!\!\!\!\!\!\!\!\!\!\!\!\!\!\!
+\, V_H({\bf r}) 
\left.\rule{0cm}{0.5cm}\right]
\varphi_i({\bf r},\eta) 
+\, \sum_{\eta'}{
\left[\rule{0cm}{0.5cm}\right.
\frac{\lambda_D}{\hbar} 
\left(\, P_x\sigma_x-P_y\sigma_y\,\right)_{\eta\eta'}}
\nonumber\\
&&
\!\!\!\!\!\!\!\!\!\!\!\!\!\!\!
+ V_{{\em xc},\eta\eta'}({\bf r})
\left.\rule{0cm}{0.5cm}\right]
\varphi_i({\bf r},\eta')
= \varepsilon_i\, \varphi_i({\bf r},\eta) \; .
\end{eqnarray}

We have solved the set of Eqs.\ (\ref{KS}) by discretizing the $xy$ plane
in a uniform grid of points and applying an iterative 
scheme to reach full selfconsistency in $V_H({\bf r})$ and 
$V_{{\em xc},\eta\eta'}({\bf r})$. 
In some cases this procedure might get trapped in a local minimum. 
Therefore, several calculations with 
different random initial conditions have to be used to ensure that the 
proper energy minimum is reached. The stability with the number of 
mesh points has also been checked. 

Figure 2 displays the density 
$\rho({\bf r})$
and spin magnetization ${\bf m}({\bf r})$
for $N=9$ electrons in a circular confining potential
with $\omega_x=\omega_y=6$ meV. We have selected this electron number
as a representative case where to check the robustness of the 
analytically predicted spin inversions. 
A SO coupling with 
$z_0=62$ {\AA} and $B=2.5$ T have been assumed. 
The magnetization density indicates the local orientation of the
spin vector and it is related to the spin-density matrix by
$m_x=2\, {\rm Re}[\rho_{\uparrow\downarrow}]$, 
$m_y=2\, {\rm Im}[\rho_{\uparrow\downarrow}]$ and
$m_z=\rho_{\uparrow\uparrow}-\rho_{\downarrow\downarrow}$.
We note from Fig.\ 2 that circular symmetry 
is conserved by both $\rho$ and $m_z$, while the horizontal 
magnetization ${\bf m}_\parallel\equiv(m_x,m_y)$ shows an angular 
dependent texture.
This result is in good agreement with 
the analytical solution given above, that predicts 
${\bf m}_\parallel({\bf r})\sim \rho(r)\; (y,x)$. 
Note also that for this particular $z_0$ and $B$ 
the vertical spin is predominantly inverted, giving a negative 
value for the total vertical spin $\langle S_z\rangle$. 
   
Figure 3 shows $\langle S_z\rangle$ as a function of $B$ and the 
intensity of the SO coupling, given by $z_0$, 
for the same dot of Fig.\ 2.
In agreement with 
the above discussion, for $B\le 2.2$ T the noninteracting model predicts 
spin inversion when decreasing the dot width. The LSDA 
also yields spin-inverted regions although with some conspicuous
differences that can be attributed to a higher rigidity in the 
electronic structure.
The interaction inhibits the spin flip at low magnetic 
fields and low widths shifting the inversion region to  
$1.8\, {\rm T} \le B \le 2.6\, {\rm T}$ and leaving only a small residue 
for $z_0\le 40$ {\AA} and $B\le 1.2$ T.
It is worth to mention that although in the 
laboratory frame $\langle S_z\rangle$ is not restricted to discrete 
values (because of the transformation $U_1$), in practice its 
fluctuations increase with the SO strength but they are generally small.
In Fig.\ 3 the deviations from $\pm\hbar/2$ when $z_0=62$ and  
48 {\AA} are  $\approx 5$ \% and 
$\approx$ 20 \%, respectively.
Experimentally the dot width can be controlled
with the electric gates; therefore, 
the results of Fig.\ 2 suggest a spin switch behavior, 
controllable with vertical magnetic and electric fields.

In Fig.\ 4 we show
the vertical spin as a function of the applied 
magnetic field and the dot deformation, for two
different values of the SO coupling. 
In this figure the mean value 
$(\omega_x+\omega_y)/2$ was kept fixed to 6 meV while 
the ratio $\delta=\omega_y/\omega_x$ was
varied to obtain different elliptical shapes.
Comparing left and right panels we see again an
interaction-induced quenching of the 
spin inversion at low magnetic fields. 
Although this occurs for the two displayed widths, at $z_0=48$ {\AA}
a larger region with inverted spin is found.
Figure 4 shows that, having fixed the SO coupling strength, 
spin inversion can be achieved in many cases either by increasing
the deformation or, alternatively, by increasing the magnetic field. 
The comparison of the $N=9$ numerical results with the analytical model
allows us to conclude that, in spite of the differences, the spin 
inversions in LSDA are qualitatively similar to the analytical ones. 
As a final piece of information we mention that the 
energy gap between the highest occupied and lowest unoccupied 
Kohn-Sham leves in the above cases stays in the range 
$[0.3,0.9]$ meV.  
This result provides a mesure of the 
ground state stability and, therefore, of the relative spin 
stiffness against thermal fluctuations.

In summary, the mechanism of spin switch through SO coupling 
has been analyzed. It has been shown that SO coupling can lead
to anticorrelated behavior in the $B$ evolution of neighboring 
levels and to up and down spin oscillations; qualitatively 
similar to experimental observations. 
The combined effects of SO coupling, weak magnetic fields and 
deformation have been studied; first analytically with a non 
interacting model and; second,
by taking into account interaction effects within the LSDA.
The diagrams with the spin dependence on these 
parameters suggest the characterization of the quantum dots
as spin switches, of relevance to spintronic technology.

This work was supported by Grant No.\ PB98-0124 from DGESeIC (Spain),
and by COFINLAB from Murst (Italy).

\begin{figure}[f]
\caption{Single particle energies in 
modified atomic units (for GaAs 1H$^*\approx 12$ meV) in the 
noninteracting model. Circular symmetry with 
$\omega_x=\omega_y=1.1$ meV
has been assumed. Each line has been shifted vertically by a small 
amount representing the charging energy. Light and dark gray color
correspond to up and down spin, respectively.}
\end{figure}

\begin{figure}[f]
\caption{ Density $\rho$ and magnetization ${\bf m}$ for a circular
dot with $N=9$ electrons, $\hbar\omega_x=\hbar\omega_y=6$ meV, 
$B=2.5$ T, and $z_0=62$ \AA. The values have been scaled by 1/9 of the 
dot central density, i.e., $\rho_0= (2.2\; 10^{-5}$~\AA$^{-2})/9$.
} 
\end{figure}

\begin{figure}[f]
\caption{Vertical spin evolution in the $B$-$z_0$ plane for the 
same quantum dot of Fig.\ 2. White (black)
color indicates upward (downward) total spin.}
\end{figure}

\begin{figure}[f]
\caption{Vertical spin evolution in the $B$-$\delta$ plane, where 
$\delta=\omega_y/\omega_x$ labels the deformation, for a fixed value
of the SO coupling. Upper row corresponds to $z_0=62$ {\AA}
in the non interacting model (left) and the LSDA (right). Lower 
row shows the same results for $z_0=48$ \AA.
We have used the same color convention of Fig.\ 3.} 
\end{figure}

\end{document}